\UseRawInputEncoding
\documentclass[aps,prb,secnumarabic, nobibnotes, twocolumn,superscriptaddress]{revtex4-1}
\usepackage{amsfonts}
\usepackage{mathrsfs}
\usepackage{amsmath}% needed for subequations
\usepackage{color}
\usepackage{natbib}
\usepackage{graphicx}
\usepackage{bm}% bold maths
\usepackage{amssymb}
\usepackage{xspace}
\usepackage{dcolumn}% Align table columns on decimal point
\usepackage{longtable}
\usepackage{siunitx} %%for angles
\usepackage{multirow}
\usepackage[colorlinks=true, letterpaper=true, pdfstartview=FitV, linkcolor=blue, citecolor=blue, urlcolor=blue]{hyperref}
\usepackage{footmisc} %%for footnote

\makeatletter

\newcommand{\Rmnum}[1]{\expandafter\@slowromancap\romannumeral #1@}
\makeatother

\begin{document}

\title{Two-dimensional Dirac semiconductor and its material realization}
\author{Botao Fu}
%\email[]{fubotao2008@gmail.com}
\affiliation{College of Physics and Electronic Engineering, Center for Computational Sciences, Sichuan Normal University, Chengdu, 610068, China}
\author{Chao He}
%\email[]{1414962823@qq.com}
\affiliation{College of Physics and Electronic Engineering, Center for Computational Sciences, Sichuan Normal University, Chengdu, 610068, China}

\author{Da-Shuai Ma}
\affiliation{Key Lab of advanced optoelectronic quantum architecture and measurement (MOE), Beijing Key Lab of Nanophotonics $\&$ Ultrafine Optoelectronic Systems, and School of Physics, Beijing Institute of Technology, Beijing 100081, China}
\author{Zhi-Ming Yu}
%\email{ygyao@bit.edu.cn}
\affiliation{Key Lab of advanced optoelectronic quantum architecture and measurement (MOE), Beijing Key Lab of Nanophotonics $\&$ Ultrafine Optoelectronic Systems, and School of Physics, Beijing Institute of Technology, Beijing 100081, China}

\author{Yong-Hong Zhao}
\affiliation{College of Physics and Electronic Engineering, Center for Computational Sciences, Sichuan Normal University, Chengdu, 610068, China}
\author{Yugui Yao}
\email{ygyao@bit.edu.cn}
\affiliation{Key Lab of advanced optoelectronic quantum architecture and measurement (MOE), Beijing Key Lab of Nanophotonics $\&$ Ultrafine Optoelectronic Systems, and School of Physics, Beijing Institute of Technology, Beijing 100081, China}

\date{\today}

\begin{abstract}
We propose a new concept of two-dimensional (2D) Dirac semiconductor which is characterized by the emergence of fourfold degenerate band crossings near the band edge and provide a generic approach to realize this novel semiconductor in the community of material science. Based on the first-principle calculations and symmetry analysis, we discover recently synthesised triple-layer (TL) BiOS$_2$ is such Dirac semiconductor that features Dirac cone at X/Y point, protected by nonsymmorphic symmetry. Due to sandwich-like structure, each Dirac fermion in TL-BiOS$_2$ can be regarded as a combination of two Weyl fermions with opposite chiralities, degenerate in momentum-energy space but separated in real space. Such Dirac semiconductor carries layer-dependent helical spin textures that never been reported before. Moreover, novel topological phase transitions are flexibly achieved in TL-BiOS$_2$: (i) an vertical electric field can drive it into Weyl semiconductor with switchable spin polarization direction, (ii) an extensive strain is able to generate ferroelectric polarization and actuate it into Weyl nodal ring around X point and into another type of four-fold degenerate point at Y point. Our work extends the Dirac fermion into semiconductor systems and provides a promising avenue to integrate spintronics and optoelectronics in topological materials.
\end{abstract}

\maketitle
%\section{Introduction}
\textit{Introduction.---}
Dirac fermion with linearly band dispersion was first discovered in 2D graphene\cite{neto2009electronic}.
Whereafter, it was rapidly popularized to 3D, and abundant topological semimetal phases including Weyl, Dirac and nodal-line semimetals are harvested in plenty of materials\cite{Yantpological2017,FangNL2016,Wengtoprev2016,Gaohengrev2019,ArmitageRevModPhys.90.015001}.
The unique electronic structures of topological semimetals lead to protected surface states and novel response to external fields and thus attract intensive research interests\cite{Wangrev2018,Nagaosa-rew-TSM}.
Actually, these quasi-fermions can be generalized towards superconductor and metallic systems\cite{WangMXNLPhysRevB.93.020503,Chiu2014PhysRevB.90.205136}.
For instance, the nodal line fermion was observed on superconducting material PbTaSe$_2$\cite{Biantpnc2016}, and the flat surface sates of nodal-line was reported widely exist in alkaline-earth metals and noble metals\cite{LiTNL2016PhysRevLett.117.096401,YanNCNL2015}.

In comparison with metal and semimetal, the semiconductor materials are particularly suitable for Dirac devices due to its high tunability and compatibility with modern electronic industry. Therefore, the introduction of Weyl/Dirac fermion into semiconductors to develop ¡°Weyl/Dirac semiconductors¡± could create a new degree of freedom for the future design of semiconductor electronic and optoelectronic devices.
Recently, the chiral Weyl fermion is theoretically predicted\cite{Hiraya2015PhysRevLett.114.206401} and experimentally observed in elemental tellurium, which is an intrinsic semiconductor with band gap about 0.36 eV, thus was dubbed as Weyl semiconductor\cite{PSakano2020hysRevLett.124.136404,PNASIdeue25530,Nanz2019}.
Based on these considerations, in this paper we will generalize the four-fold degenerating Dirac fermion from semimetal to semiconductor in two-dimension. Because 2D materials possess prior mechanical properties and small size that are more favorable for integration and regulation.

\begin{figure}
	\includegraphics[width=8cm]{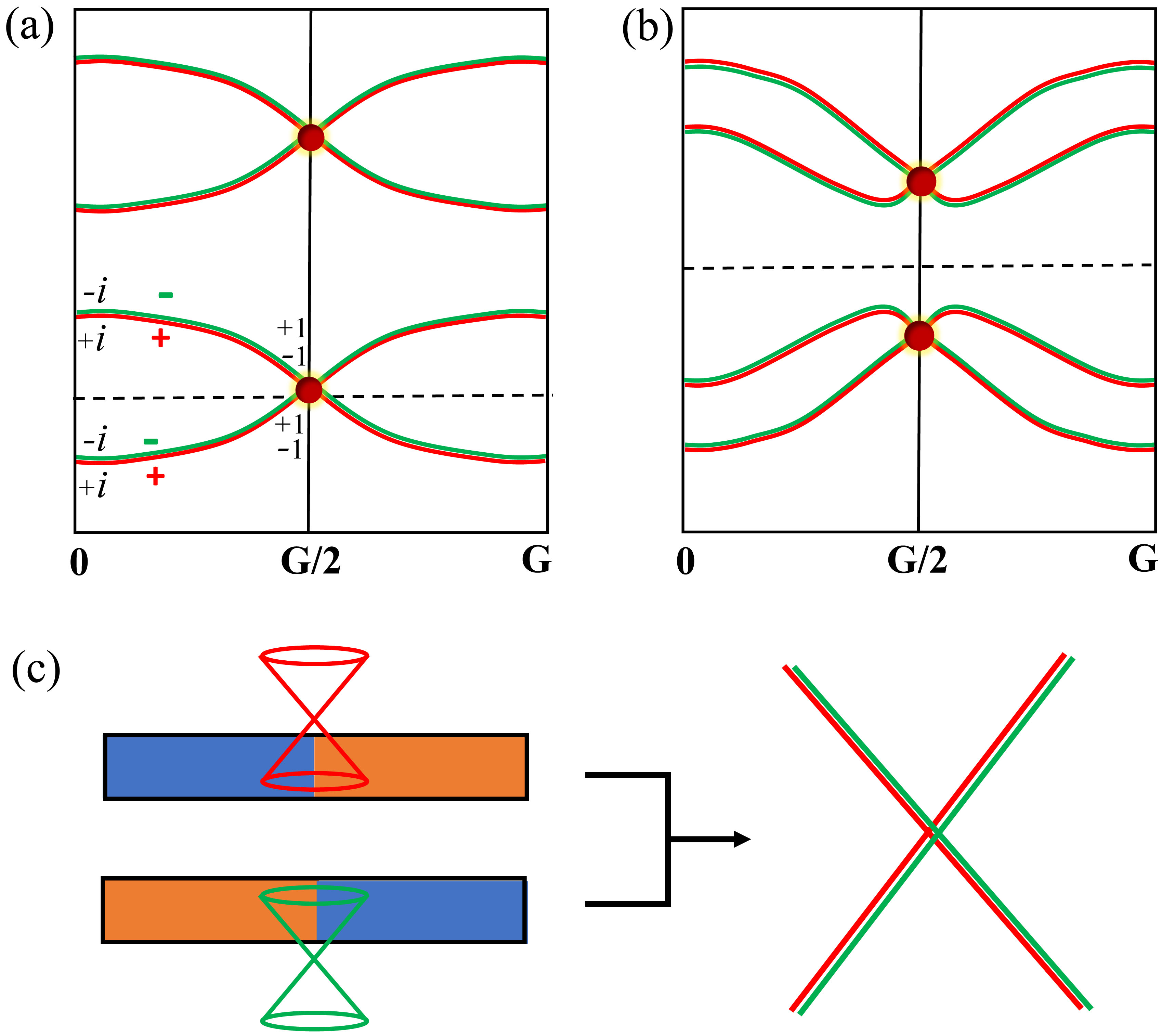}
	\centering
    \caption{A nonsymmorphic operator $\widetilde{g}=\{g|\boldsymbol t\}$ leads to band crossings at $\mathbf{G}/2$ ($\mathbf{G}$ is the reciprocal lattice vector) for a system with $\mathcal{PT}$ symmetry in the present of SOC. (a) The electronic structure of system with electron filling number is $4n+2$. (b) The electronic structure of system with electron filling number is $4n$. The dash line indicates the Fermi level.
    (c) Schematic illustration for construction of Dirac fermion in multiple-layer structure. Two Weyl fermion locate on individual monolayers merge together to form a Dirac fermion.
    }
    \label{Fig1}
\end{figure}
\textit{Proposal of 2D Dirac semiconductors---}
Firstly, we review the concept of 2D Dirac semimetal in the presence of spin-orbital coupling (SOC), as proposed by S. M. Young at al\cite{Young2DDSM2015}.  For a system with both time reverse $\mathcal{T}$ and space inversion $\mathcal{P}$ in addition with a nonsmmorphic symmetry $\widetilde{g}=\{g|\boldsymbol t\}$, where $g$ is point group operator and $\boldsymbol t$ is fractional translation.
As shown in Fig. \ref{Fig1}(a), along a $g$-invariant line, e.g. $\mathbf{0}$ to $\mathbf{G}$, each double-degenerated band can be marked by the opposite and ${\boldsymbol{k}}$-dependent eigenvalue of $\widetilde{g}$: $\pm i e^{i\boldsymbol {kt}}$.
It equals to $\pm i$ at $\mathbf{0}$ and gradually evolves into $\pm 1$ at $\mathbf{G}/2$.
At $\mathbf{G}/2$, the $\mathcal{T}^2=-1$ guarantees two $+1$ states or $-1$ states to be degenerate, namely Kramers degeneracy.
Consequently, two branches of double-degenerate bands have to stick together and form a four-fold degenerate Dirac point at $\mathbf{G}/2$.
If the electron filling number satisfies $4n+2 (n\geq 0)$, the Fermi level then crosses Dirac point and it becomes an ideal Dirac semimetal with point-like Fermi surface.

Despite concise and intriguing picture, the practical and ideal 2D Dirac semimetal materials are very rare\cite{Guans2017PhysRevM,Kowalcyz2020}.
One reason may be the limitation of electron filling number. By counting the 2D nonsmmorphic materials in material database\cite{2018Twomounet}, we find only 16$\%$ of them have $4n+2$ electron filling number while the rest host $4n$ electrons.
The other reason may be that the second-order SOC effect is relatively small compared with second-order electron hopping which results in the emergence of undesirable trivial metal states at Fermi level. This can be explicitly verified through a simple tight-binding model\cite{Supplefbt}.
From other perspective, as in Fig. \ref{Fig1}(b) for most materials with $4n$ electrons, the Fermi level lies inside the gap between two sets of bands including Dirac fermions at $\mathbf{G}/2$. In general, the system is a semiconductor.
Suppose the position of band edge happens to reside at or nearby $\mathbf{G}/2$, it will become so-called ``Dirac semiconductor" with symmetry-enforced Dirac cone close to Fermi level. In such Dirac semiconductor, the signature of Dirac fermions could be probed by angle-resolved photoemission spectroscopy (ARPES) and transport experiments as what have been done in Weyl semiconductors\cite{Kowalcyz2020,Nanz2019,QiuGNt2020}.

In Fig. \ref{Fig1}(c), we introduce an practical approach to construct this Dirac semiconductor in multiple-layer systems.
Suppose a monolayer system without $\mathcal{P}$ but with $\mathcal{T}$ symmetry,
with SOC being taken into consideration each band becomes spin splitting
except for that at the time-reversal invariant momenta (TRIM) where the Kramers degeneracy occurs.
It's worth noting that such Kramers degeneracy is recently reported host unanticipated topological charge in 3D chiral lattice, noted as Kramers-Weyl fermion\cite{CahngGNM2018,LiGNC2019}.
Bearing this in mind, we stack up two $\mathcal{P}$-breaking monolayers which are semiconductor with band edge at $\mathbf{G}/2$, and meanwhile impose an in-plane fractional translation. The constructed bilayer system hosts both $\mathcal{P}$ and glide mirror symmetries.
Consequently, two Kramers-Weyl points from two monolayers will merge one Dirac point at $\mathbf{G}/2$.
This tells us that Dirac semiconductor can exist in certain stacked bilayer or some intrinsic bilayer or multiple layer materials.
More importantly, with unique multiple-layer structure and semiconducting nature, the Dirac semiconductor will exhibit flexibly tunable electronic and topological properties under external fields.

\begin{figure}
	\includegraphics[width=8.5cm]{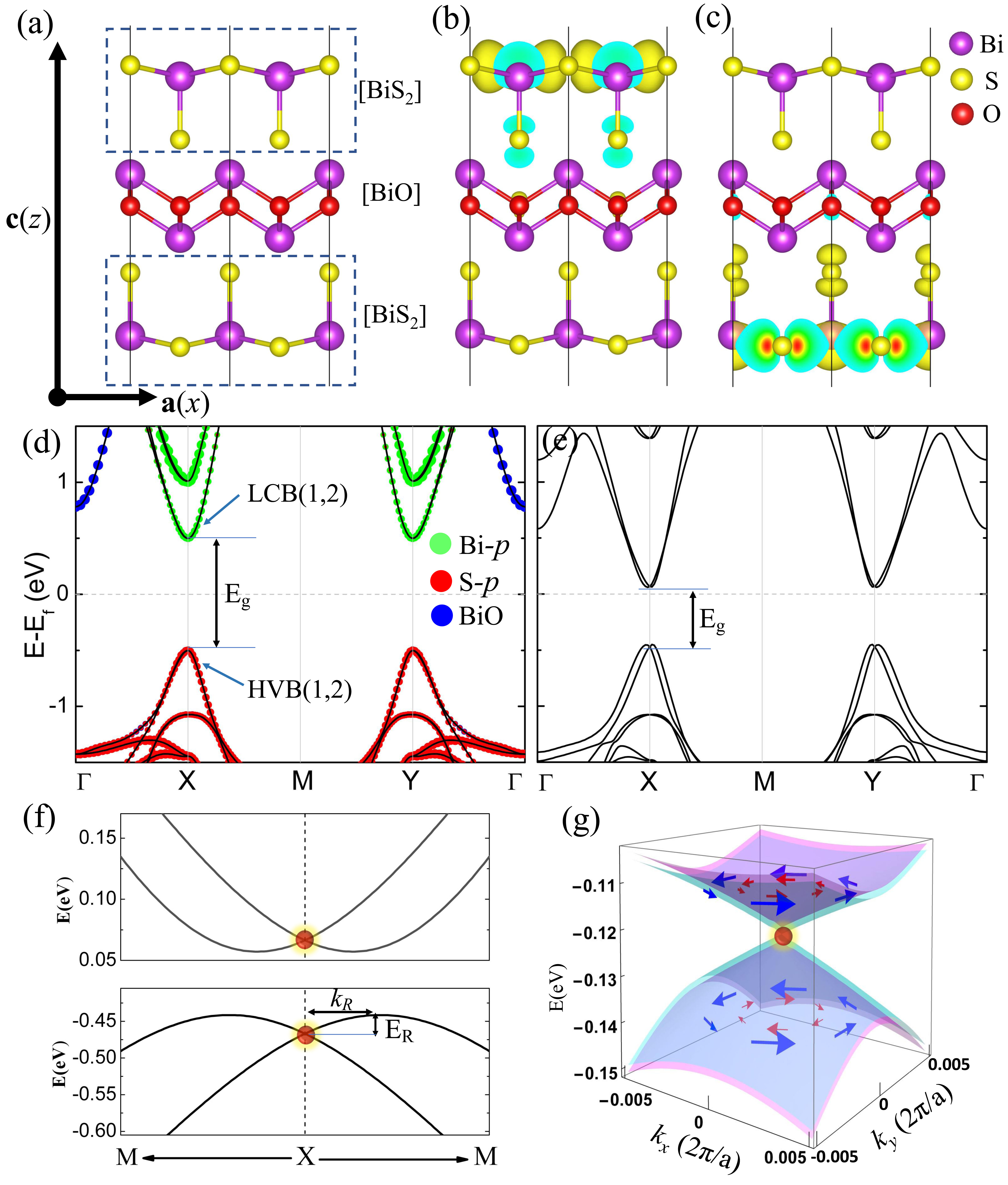}
    \caption{(a) The atomic structure of TL-Bi$_2$OS$_2$. (b), (c)The real-space charge density distribution of the maxima of HVB1 and HVB2. (d) The orbital-resolved band structure without SOC effect. The green color represents Bi's $p$-orbital and the red color represents S's $p$-orbital on BiS$_2$ layer, while the blue color represents all orbitals from BiO layer. (e) The band structure with SOC effect. (f) The enlarged figures around LCB and HVB with SOC. (g) The 3D Dirac cone around $X$ with helical spin textures indicated by arrows. The different color represent bands from different BiS$_2$ layers. The Dirac fermions are marked by yellow dots.}\label{Fig2}
\end{figure}

\textit{Material Realization---}
Based on above analysis, we find out four kinds of multiple-layer semiconductors\cite{Supplefbt} which host nonsmmorphyic space group and proper band edge position. Here, taking  TL-BiOS$_2$ as a prototype, we will demonstrate the emergence of Dirac fermion around the band edge and its manipulation by external fields. The TL-BiOS$_2$ is predicted to be a stable direct band-gap semiconductor with high carrier mobility and large visible light absorption\cite{MaterilsZhangxw2018}.
In experiment, the high carrier mobility as well as power conversion effciency as solar cell has been recently observed in 2D BiOS$_2$\cite{Huang2DM_2020}.
As shown in Fig. \ref{Fig2}(a), the TL-BiOS$_2$ consists of one BiO layer in the middle and two BiS$_2$ layers on the top and bottom, forming sandwich-like structure. The optimized lattice constance is $|\boldsymbol{a}|$=$|\boldsymbol{b}|$=3.95 \AA.
Although, each BiS$_2$ layer is inversion-asymmetric with non-centrosymmetric polar site point group $C_{4v}$. Two BiS$_2$ layers become each other's inversion partners and meanwhile have inter-layer relative translation, $(\boldsymbol{a}+\boldsymbol{b})/2$.
Hence the whole system hosts centrosymmetric and nonsymmorphic space group $P4/nmm$ (No.129).

The electronic structures of TL-Bi$_2$OS$_2$ under generalized gradient approximation (GGA) are displayed in Fig. \ref{Fig2}(d)-(e).
In the absence of SOC, it has a direct band gap ($E_g$=0.97 eV) with both valence band
maximum (VBM) and conduction band minimum (CBM) located at X/Y point.
The quivalentence of X and Y is guaranteed by the $C_{4z}$ symmetry.
Importantly, both the highest valence band (HVB) and lowest conduction band (LCB) are double-degenerate not only at X/Y point but along X-M-Y path.
This degeneracy is protected by screw axis $\widetilde{C}_{2x}$/$\widetilde{C}_{2y}$\cite{Fanmeprb2018}, in which the tilde refers to additional $(\boldsymbol{a}+\boldsymbol{b})/2$ translation.
The orbital-resolved band structure demonstrates that both HVB and LCB primarily derive from BiS$_2$ layers, but the former from S-$p$ orbital while the latter from Bi-$p$ orbital.
The charge density distributions of double-degenerate HVB (HVB1 and HVB2) states are calculated in Fig. \ref{Fig2}(b)-(c), from which we find despite their degeneracy in momentum-energy space, the HVB1 and HVB2 locate separately on top and bottom BiS$_2$ layers, respectively.
Similar phenomenon also happens for the LCB.
Thus, TL-Bi$_2$OS$_2$ well satisfies the conditions for the emergence of Dirac semiconductor: (i) Two $\mathcal{P}$-breaking BiS$_2$ layers stack up via electrostatic interaction with intermediate BiO$_2$ layer which simultaneously host fractional translation and inversion symmetries.
(ii) The double-degenerate VBM or CBM at X/Y in TL-Bi$_2$OS$_2$ originate from opposite BiS$_2$ layers.
%%%%%%%%%%%%%%%

When considering SOC effect in Fig. \ref{Fig2}(e), each band becomes double degenerate because of $(\mathcal{PT})^2=-1$.
The band gap ($E_g$) at X/Y reduces from 0.97 eV to 0.40 eV.
More interesting, the LCB and HVB split along XM path and exhibits Rashba-like dispersions as demonstrated in Fig. \ref{Fig2}(f).
The Rashba parameter can be defined as $\alpha_R=2E_R/k_R$, which is about 1.27 {eV\AA} and 2.15 {eV\AA} for HCB and LVB, respectively. In reality, it's distinctive from traditional Rashba splitting which occurs in non-centrosymmetric materials\cite{ishizaka2011giant,maass2016spin,ZhangshPhysRevB.100.165429,di2013electric}.
Around X/Y point, two double-degenerated bands cross each other to form a fourfold-degenerated Dirac fermion as expected.
This could be understood from following perspective.
Since the HVB1 and HVB2 locate separately on opposite BiS$_2$ layers without $\mathcal{P}$, the SOC effect certainly gives a Rashba splitting for them.
Then considering two BiS$_2$ layers in TL-Bi$_2$OS$_2$ as a whole, the nonsmmorphic operation in combination with space inversion will stick two sets of Rashba bands together, and in particular enables the appearance of Dirac fermion at X/Y.
Therefore, although this Dirac semiconductor host $\mathcal{PT}$ symmetry, one can identify the spin textures of Dirac fermion.
As shown in Fig. \ref{Fig2}(g), it indeed hosts opposite chiral spin textures on different BiS$_2$ layers, which may provide potential applications in spintronics.

\begin{figure*}
	\includegraphics[width=17.0cm]{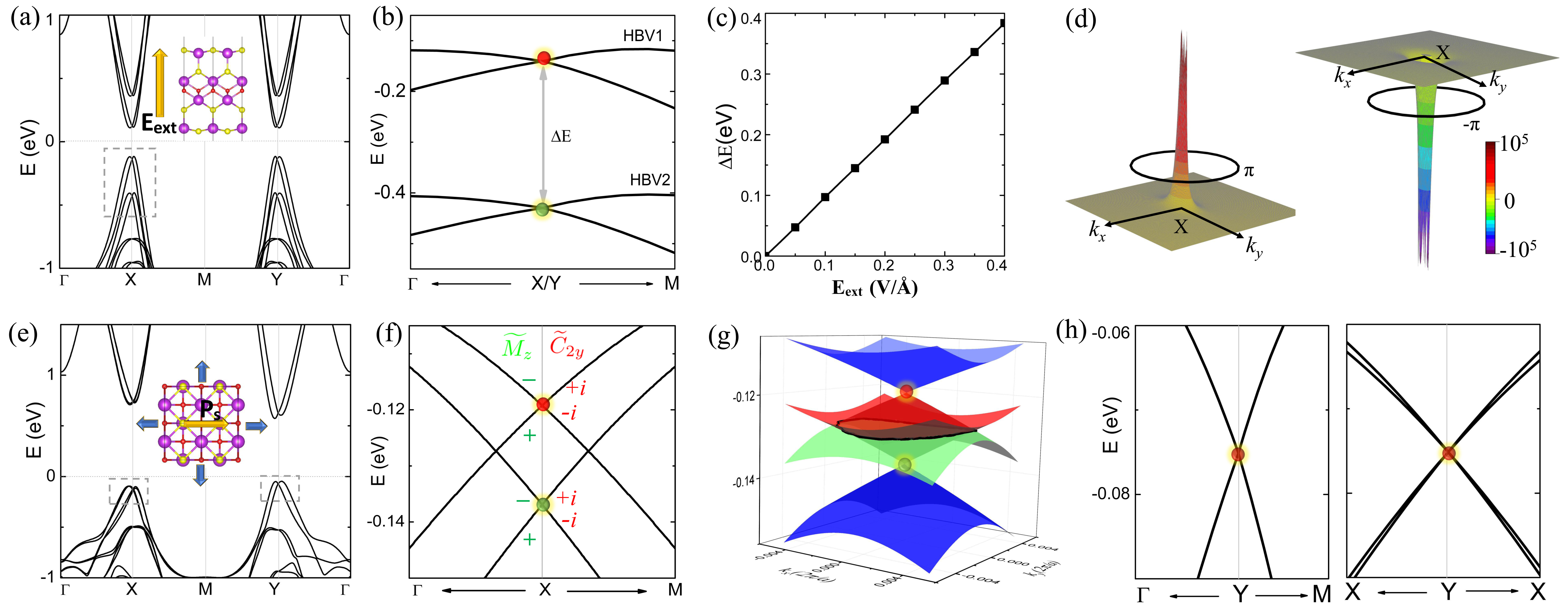}
    \caption{ (a) The electronic structure under external vertical electric filed, E$_{\text{ext}}$=0.3 V/\AA. (b) The enlarged view around valence band. (c) The relation of $\Delta E$ and external electric filed. (d) The distribution of Berry curvature around two Weyl points at X, and the Berry phases circled them are given. (e) The electronic structure under $2\%$ biaxial strain, with spontaneous electric polarization $P_s$ along $y$. (f) The enlarged view around the valence band at X. The eigenvalues of $\widetilde{M}_{z}$ and $\widetilde{C}_{2y}$ are given. (g) The 3D view of bands at (f), the nodal line is labelled by black line. (h) The enlarged views around Y along different $k-$paths. The red and green dot indicates the different Weyl fermions at X in (b) and (f), the Yellow dot represents Dirac fermion in (h). }\label{Fig3}
\end{figure*}
\vspace{12pt}

To capture the physics of Dirac fermion around band edge, we are going to build an effective $k{\cdot}p$ model.
For the VBM at X point, we can choose the degenerate HVB1 and HBV2 states as the basis in the absence of SOC.
Then considering SOC effect the Hamiltonian around X is written as,
\begin{equation} \label{hlt1}      %
H=
\left(                 %
  \begin{array}{cc}   %
     h_{+} & T \\  %
     T &  h_{-}\\  %
  \end{array}
\right),                 %
\end{equation}
\begin{equation}\label{hlt2}
h_{\pm}=\epsilon_{k} \pm(\alpha_{y}k_y\sigma_x-\alpha_{x}k_x\sigma_y ),
\end{equation}
\begin{equation}\label{hlt3}
\epsilon_{k}= \frac{\hbar^{2}k_x^2}{2m_x^{*}}+\frac{\hbar^{2}k_y^2}{2m_y^{*}},
\end{equation}
where $h_{\pm}$ describes the electron from HVB1/HVB2 that locates on top/bottom BiS$_2$ layer and $T$ is the inter-layer coupling term.
The $\alpha_{x,y}$ and $m_{x,y}^{*}$ are the anisotropic Rashba parameters and effective mass along $x/y$ directions, respectively.
The $\sigma_{x,y,z}$ is Pauli matrix that represents spin degree of freedom.
In Eq. \ref{hlt2} we consider the predominant Rashba SOC term that originated from the interfacial polar filed between BiO and BiS$_2$ layer.
Since the top and bottom BiS$_2$ layers feel opposite polar filed, they possess opposite Rashba spin texture as indicated on $h_{\pm}$.
At X point the little group is $D_{2h}$, the time reverse and space inversion are chosen as $\mathcal{T}=i\sigma_y K \tau_z$ and $\mathcal{P}=\sigma_0 \tau_y$, then based on specific commutation relations other symmetric operations can be readily written as
$\widetilde{C}_{2x}=\sigma_x \tau_x$, $\widetilde{C}_{2y}=i\sigma_y \tau_y$ and $\widetilde{M}_{z}=\sigma_z\tau_x$, where
the $\tau_{x,y,z}$ is Pauli matrix that represents layer degree of freedom.
With these constraints the inter-layer term is given as $T=t_i k_x \sigma_0\tau_x$, $t_i$ is inter-layer coupling coefficient.
Hereafter, the low-energy effective Hamiltonian at X with first-order approximation is obtained as
\begin{equation}\label{hlt4}
{H_\text{X}=(\alpha_{y}k_y\sigma_x-\alpha_{x}k_x\sigma_y )\tau_z+t_ik_x\sigma_0\tau_x}.
\end{equation}
With a straightforward solving, we get two branches of doubly degenerate energy spectra,
\begin{equation}\label{hlt5}
{E_{\pm}^\text{X}=\pm \sqrt{(\alpha_x^{2}+t_i^{2})k_x^{2}+\alpha_y^{2} k_y^{2}}},
\end{equation}
which indeed depicts a quadruple degenerated Dirac fermion at $k_x$=$k_y$=0 with Fermi velocities $v_x=\sqrt{\alpha_x^{2}+t_i^{2}}$ and $v_y=|a_y|$.
By fitting the $k\cdot p$ model with our first-principle calculations, we obtain $\alpha_x$=1.40 eV\AA, $\alpha_y$=2.15 eV\AA, $t_i$=0.09 eV\AA.
The $t_i$ is much smaller than $\alpha_{x,y}$ indicates the interlayer interaction is weak and can be reasonably neglected.
Then the Dirac fermion in Eq. \ref{hlt4} is able to be decomposed into two Weyl fermions with opposite layer index,
\begin{equation}\label{hlt6}
H_{\pm}^W=\pm(\alpha_y k_y\sigma_x-\alpha_xk_x\sigma_y).
\end{equation}
It indicates the Dirac fermion can be approximately taken as two spatially isolated Weyl fermions with opposite layer-index and carrying opposite helical spin texture, which differs from other known Dirac fermion in semimetal or metal systems nd may extend the potential application of Dirac fermion.
For instance, the linearly dispersion as well as in-plane helical spin texture in Dirac semiconductor may suppress direct backscattering and facilitate carrier mobility, and the layer-resolved spin polarization\cite{zhang2014hidden,QHliua2015,QHliuPRB2016,GSPhysRevB} can be flexibly controlled under external filed.
Moreover, we find for two and three layers of TL-BiOS$_2$, the Dirac fermion still exists\cite{Supplefbt}, making it easy to probe in experiment. When extending to 3D bulk limitation, the Dirac fermion transforms into a Dirac nodal line along $k_z$ direction, where each point can be taken as a Dirac point, thus dubbed as 3D Dirac nodal-line semiconductor.
We believe these unique features in such Dirac semiconductors deserve further experimental explorations.

\textit{Manipulation of TPTs.---}

The topological phase transition (TPT) manipulated by external electric field is of great significance, especially in lower dimension\cite{li2009electric,LicH2017PhysRevB,liu2015switching}.
For example, the electric-field-tuned TPT is recently reported in ultra-thin Na$_3$Bi\cite{Nature2020}, a well-known Dirac semimetal.
Since Dirac fermion is protected by $\mathcal{PT}$ in addition with other crystal symmetries, breaking either $\mathcal{P}$ or $\mathcal{T}$ will induce various TPTs\cite{Young2DDSM2015}.
The fact that the Dirac fermion in TL-BiOS$_2$ contains two spatially isolated Weyl points provides a good opportunity to manipulating by vertical electric field.
Also, the semiconducting nature of TL-BiOS$_2$ is in favor of electric field control than semimetal or metal systems in which the electrostatic screening predominates.
Specifically, an vertical electric field breaks inversion symmetry of TL-BiOS$_2$ by inducing stacking potential on different BiS$_2$ layers. This process can be approximately depicted by adding a mass term $V_1=V_z\sigma_0\tau_z$ to $H_\text{X}$.
Neglecting interlayer coupling term, $H_\text{X}+V_1$ resolves into ${V_z\sigma_0 + H_{+}^W}$ and ${-V_z\sigma_0 + H_{-}^W}$, which indicate a Dirac fermion can be successful split into two Weyl points with energy difference $\Delta E=2V_z$.
In Fig. \ref{Fig3} (a)-(c), we perform first-principle calculations with external vertical electric filed (E$_{\text{ext}}$) varying from 0.0 to 0.4 V/\AA.
We find the four-fold degenerated Dirac fermions at HVB and LCB at X/Y split into a pair of Karamer-Weyl fermions.
The split amplitude $\Delta E$ is linearly proportional to the size of E$_{\text{ext}}$ with a ratio of 0.96 e{\AA} as shown in Fig. \ref{Fig3}(c).
We already know these two Weyl fermions have layer-dependent chiral spin textures.
Moreover, we show they also demonstrate large Berry curvature distributions and can be identified by the quantized $\pm \pi$ Berry phase surrounding them as displayed in Fig. \ref{Fig3}(d).
Hence by switching Electric field direction, the spin polarization direction and spin Hall coefficient can be modified, working as spin filed effect transistor (SFET)\cite{NLliu2013tunable}.

Strain is another effective method for engineering the electronic, magnetic and topological properties of 2D materials\cite{Xucz2017PRL,ZhaoCX2020PRL}.
As we know the Dirac point in TL-Bi$_2$OS$_2$ is guaranteed by unique nonsymmorphic space group, so called symmetry-enforced that robust against any symmetry-remained perturbations. In principle, a simple uniaxial or biaxial strain on TL-Bi$_2$OS$_2$ won't change the nature of nonsymmorphic.
%Luckily, we find TL-Bi$_2$OS$_2$ a piezoelectric material\cite{cui2018two,guan2020recent} with piezoelectric coefficient of {\color{red}$XXX5.0 P/cm$}. %{\color{red}P$_s$= 15$\mu$C/cm$^2$}
Luckily, we find TL-Bi$_2$OS$_2$ is a novel piezoelectric material\cite{cui2018two,guan2020recent}. A very tiny biaxial extensive strain will induce
a Peierls-like atomic distortion on BiS$_2$ layers along $y$ direction that breaks space inversion and gives switchable in-plane electronic polarization\cite{Supplefbt}.
In Fig. \ref{Fig3}(e), we calculate the  bandstructure under $2\%$ biaxial strain.
One can find the band gap at X point (0.85 eV) becomes larger than that at Y (0.70 eV). Because the in-plane polar distortion breaks the $C_{4z}$ rotation and lifts the inter-valley degeneracy between X and Y.

The space group of ferroelectric distorted TL-BiOS$_2$ is $Pmn2_1$, which contains four symmetry operations: $M_x$, $\widetilde{M}_{z}$, $\widetilde{C}_{2y}$ and identity.
At X point, the lacking of $\mathcal{P}$ and $\widetilde{C}_{2x}$ permits the emergence of a mass term $V_2=m_0\sigma_z\tau_x$, which couples two BiS$_2$-layers and opens a gap. As demonstrated in Fig. \ref{Fig3}(f), the Dirac fermion splits into two Karamer-Weyl fermions at X point with gap $2m_0$=18.1 meV.
Moreover, thanks to surviving $\widetilde{M}_{z}$, all bands could be characterized by its eigenvalues within the whole Brillouin zone.
We find along X$\Gamma$ path two bands with opposite $\widetilde{M}_{z}$ eigenvalues have to cross each other and exhibits unique hourglass-like band connection. Actually, such band crossing happens for arbitrary $k$-path connecting X and $\Gamma$ and forms a closed nodal ring rather than isolated points.
To confirm this, in Fig. \ref{Fig3}(g) we draw the 3D band spectrum and find a Weyl nodal ring surrounding X appears in accompany with two Weyl points at X, in consistent with our symmetry analysis.
Besides, the band crossing along XM is also protected by distinctive eigenvalues of $\widetilde{C}_{2y}$.
So far as we know such glide mirror protected hourglass nodal rings or nodal chains are mainly studied on 3D systems\cite{Bzduseknature2016}, our result provides a practical and tunable Weyl nodal ring candidate in lower dimension.
At Y point, despite sharing same little group as X point, it hosts distinctive commutation relations, the zero-order mass term $m_0\sigma_z\tau_x$ is forbidden by $\widetilde{C}_{2y}=\sigma_y \tau_x$.
Therefore, the quadruple degeneracy is maintained at Y point while the $\mathcal{P}$-breaking allows first-order term $V_3=w_yk_y\sigma_x\tau_y+w_xk_x\sigma_y\tau_y$, which gives linear band splitting apart from Y along general $k-$path e.g. Y-X-Y as shown in Fig. \ref{Fig3}(h).
Note the double band degeneracy are still maintained along $\Gamma$Y and YM which are protected by the anticommutation relation $\{\widetilde{M}_{x}, \widetilde{C}_{2y}\}$=0 and by $\mathcal{T}\widetilde{C}_{2y}=-1$, respectively.
This type of Dirac point that even exists without $\mathcal{PT}$ is very recently proposed in monolayer SbSSn\cite{JinDSM2020PRL}.

\textit{Conclusion.---}
As a summary, in this work, we put forward the concept of 2D Dirac semiconductor that host Dirac cone around band edge in nonsymmorphic systems and
then we propose an effective approach to search for such Dirac semiconductor in certain multiple materials.
We explicitly demonstrate an practical Dirac semiconductor materials: TL-BiOS$_2$, in which the Dirac cone is composed of two spatially isolated Weyl cones, and provides layer-resolved spin polarization. Moreover, with various TPTs can be induced in TL-BiOS$_2$ by imposing external strains or electric fields.
Besides, it is worth mentioning that starting from Dirac semiconductor, one can conveniently achieve Dirac semimetal state by simply shifting Fermi level through imposing gate voltage, element doping as well as band inversion.
In fact, the F-substitution of O is used to modify the Fermi level in bulk LaOBiS$_2$ to realize high-temperature superconducting state\cite{Yazici2015sc}, and some proposed Dirac semimetal materials like HfGeTe can be taken as Dirac semiconductor with band inversion\cite{Guans2017PhysRevM}.
Our finding of 2D Dirac semiconductor not only enriches topological quantum material families but also provides a new avenue for exploring spin polarization, ferroelectric and optoelectronic properties in topological quantum materials.

%%%%%%%%%%%%%%%%%%%%%%%%%%%%%%%%
\begin{acknowledgements}
This research was funded by the National Natural Science Foundation of China (Grant No. 11874273 and No. 11974009) and the Key Project of Sichuan Science and Technology Program (2019YFSY0044). We also thank the Sichuan Normal University for financial support (No. 341829001). This research project was supported by the High Performance Computing Center of Sichuan Normal University, China.
The work at BIT is supported by the National Key R\&D Program of China (Grant No. 2016YFA0300600), the National Natural Science Foundation of China (Grants Nos. 11734003), the Strategic Priority Research Program of Chinese Academy of Sciences (Grant No. XDB30000000).

\end{acknowledgements}

%\section*{References}
%\setcitestyle{square}
%\setcitestyle{citesep={,}}
%setcitestyle{citesep={,}}
%\bibliographystyle{achemso}
%\bibliographystyle{elsarticle-num-names}
\bibliography{paperref}

\end{document}